%% file: collaboration.tex
\pdfoutput=1

\documentclass[11pt]{article}

\usepackage{acl}

\usepackage{times}
\usepackage{latexsym}

\usepackage[T1]{fontenc}

\usepackage[utf8]{inputenc}

\usepackage{microtype}

\usepackage{inconsolata}

\usepackage{amsmath}
\usepackage{graphicx}
\usepackage{multirow}
\usepackage{tabularx}

\title{Beyond Frameworks: Unpacking Collaboration Strategies in Multi-Agent Systems}

\author{Haochun Wang, Sendong Zhao$\thanks{ \ \ Corresponding author}$, Jingbo Wang, Zewen Qiang, Bing Qin and Ting Liu
 \\
Research Center for Social Computing and Information Retrieval, \\Harbin Institute of Technology, China \\
\texttt{\{hcwang,sdzhao\}@ir.hit.edu.cn} \\
}

\begin{document}
\maketitle
\begin{abstract}
\input{Text/0.abstract}
\end{abstract}

\input{Text/1.intro}

\input{Text/2.related}
\input{Text/3.collaboration}
\input{Text/4.experiment}
\input{Text/6.conclusion_ethics}


\newpage
\clearpage
\input{Text/7.Limitations}
\bibliography{custom}


\input{Text/x.Appendix}

\end{document}

%% file: Text/0.abstract.tex
Multi-agent collaboration has emerged as a pivotal paradigm for addressing complex, distributed
tasks in large language model (LLM)-driven applications.
While prior research has focused on high-level architectural frameworks, the granular mechanisms
governing agents, critical to performance and scalability, remain underexplored.
This study systematically investigates four dimensions of collaboration strategies:
(1) agent governance, (2) participation control, (3) interaction dynamics, and (4) dialogue history management.
Through rigorous experimentation under two context-dependent scenarios—Distributed Evidence Integration (DEI) and Structured
Evidence Synthesis (SES)—we quantify the impact of these strategies on both task accuracy and computational efficiency.
Our findings reveal that centralized governance, instructor-led participation, ordered interaction
patterns, and instructor-curated context summarization collectively optimize the trade-off between
decision quality and resource utilization with the support of the proposed Token-Accuracy Ratio (TAR).
This work establishes a foundation for designing adaptive, scalable multi-agent systems,
shifting the focus from structural novelty to strategic interaction mechanics.

%% file: Text/1.intro.tex
\section{Introduction}

\begin{figure*}[htb]
    \centering
    \includegraphics[width=0.99\textwidth]{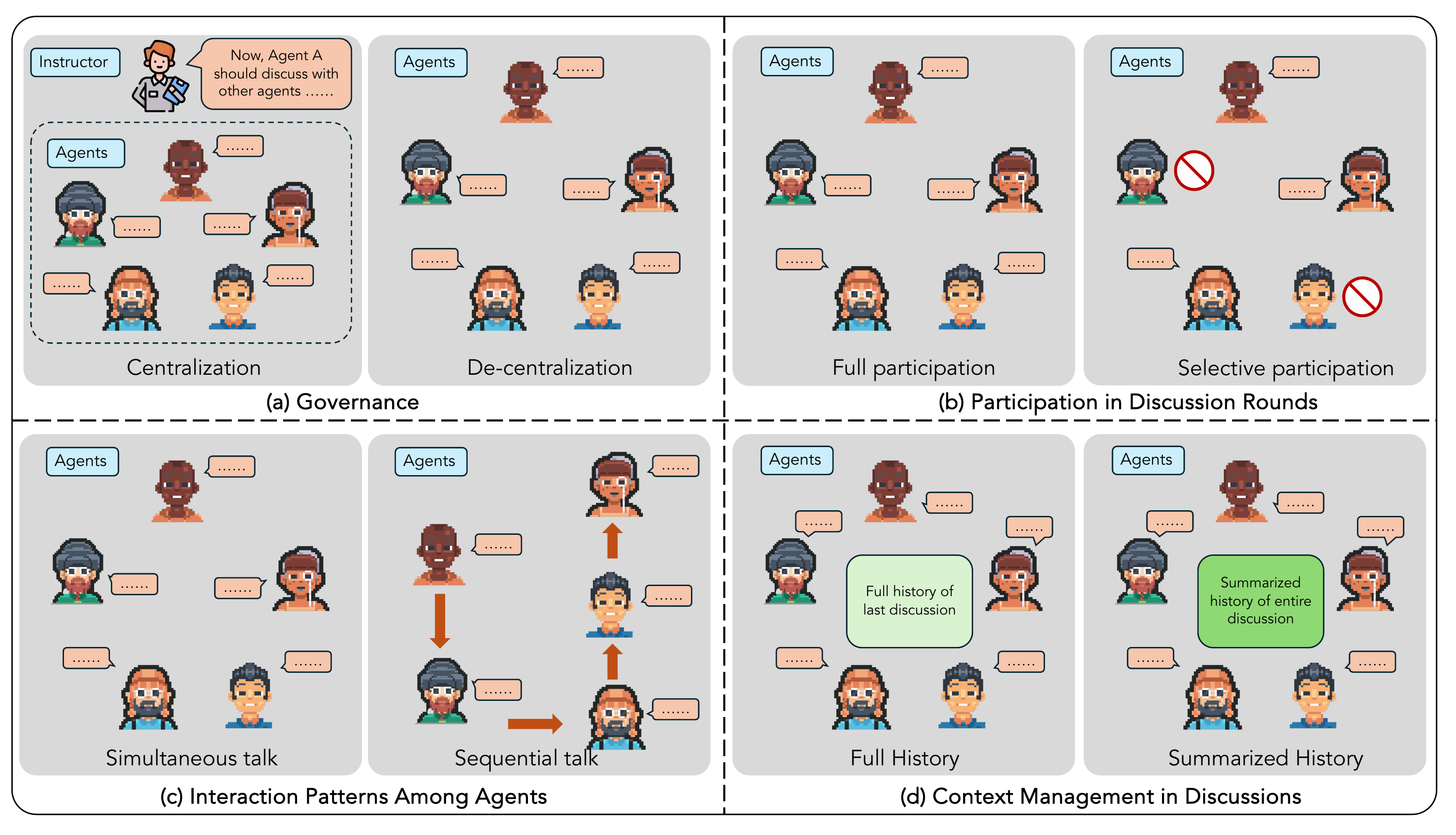}
    \caption{Illustration of collaboration strategies for multi-agent systems, including (a) Governance, (b) Participation in Discussion Rounds, (c) Interaction Patterns Among Agents, and (d) Context Management in Discussions.}
    \label{fig:intro}
\end{figure*}

The advent of large language models (LLMs) ~\cite{touvron2023llama,chatgpt,Anthropic2025Claude,zeng2023glm-130b}
has catalyzed transformative advances in autonomous reasoning and decision-making, enabling multi-agent systems
~\cite{chanchateval,chen2023agentverse,zhugegptswarm,du2024multi} to tackle tasks that exceed the cognitive or functional
limits of individual agents.
Such systems are increasingly deployed in domains ranging from healthcare diagnostics~\cite{kim2024mdagents}
to scientific discovery~\cite{su2024two}, where collaborative synthesis of specialized expertise is paramount.
However, as system complexity scales, a critical gap persists: existing frameworks prioritize structural
architectures and role assignments but neglect the granular mechanics of agent collaboration: how agents
dynamically interact, share context, and reach consensus.

Current approaches often assume rigid pipeline workflows, where agents sequentially process subtasks.
While effective for linear workflows, this paradigm fails to capture the nuanced deliberation
of human teams, where domain experts iteratively refine decisions despite individual competence.
Key questions remain unanswered, ``\textit{who speaks, when, to whom, and with what context?}'',
which includes: (a) How should agents govern their interactions?
(b) When and to whom should they communicate?
(c) How can contextual depth be balanced against computational costs?

In this study, we address these gaps by formalizing four dimensions of multi-agent collaboration:

\textit{(1) Governance.} Centralized systems use an instructor agent to coordinate interactions, ensuring decision-making. Decentralized systems allow agents to self-organize, promoting autonomy but risking coordination challenges.

\textit{(2) Participation in Discussion Rounds.} Full participation involves all agents in every round. Selective participation engages only relevant agents, optimizing efficiency but limiting perspectives.

\textit{(3) Interaction Patterns Among Agents.} Agents may broadcast to all, target-specific peers, or follow specific turns. These patterns influence information clarity, relevance, and the speed of consensus.

\textit{(4) Context Management in Discussions.} Systems either retain full dialogue history for depth or use summarization for efficiency, balancing situational awareness with computational cost.

We evaluate these strategies through extensive experiments under two context-dependent scenarios:
Distributed Evidence Integration (DEI) and Structured Evidence Synthesis (SES).
Our results demonstrate that centralized governance, coupled with instructor-led participation and context management,
reduces token costs by up to 93.0\% while maintaining even better accuracy.
Conversely, decentralized systems exhibit higher variance and computational inefficiency, particularly in SES, where misaligned agents degrade performance.
The introduced Token-Accuracy Ratio (TAR) further quantifies the trade-offs, guiding practitioners toward resource-efficient configurations.
By bridging the gap between high-level architectural design and low-level interaction mechanics, this work advances
the development of adaptive, context-dependent multi-agent systems.
It underscores the necessity of strategic collaboration protocols in scaling LLM-based
applications, offering actionable guidelines for future research in dynamic and real-world environments.

%% file: Text/2.related.tex
\begin{figure*}[tb]
    \centering
    \includegraphics[width=0.91\textwidth]{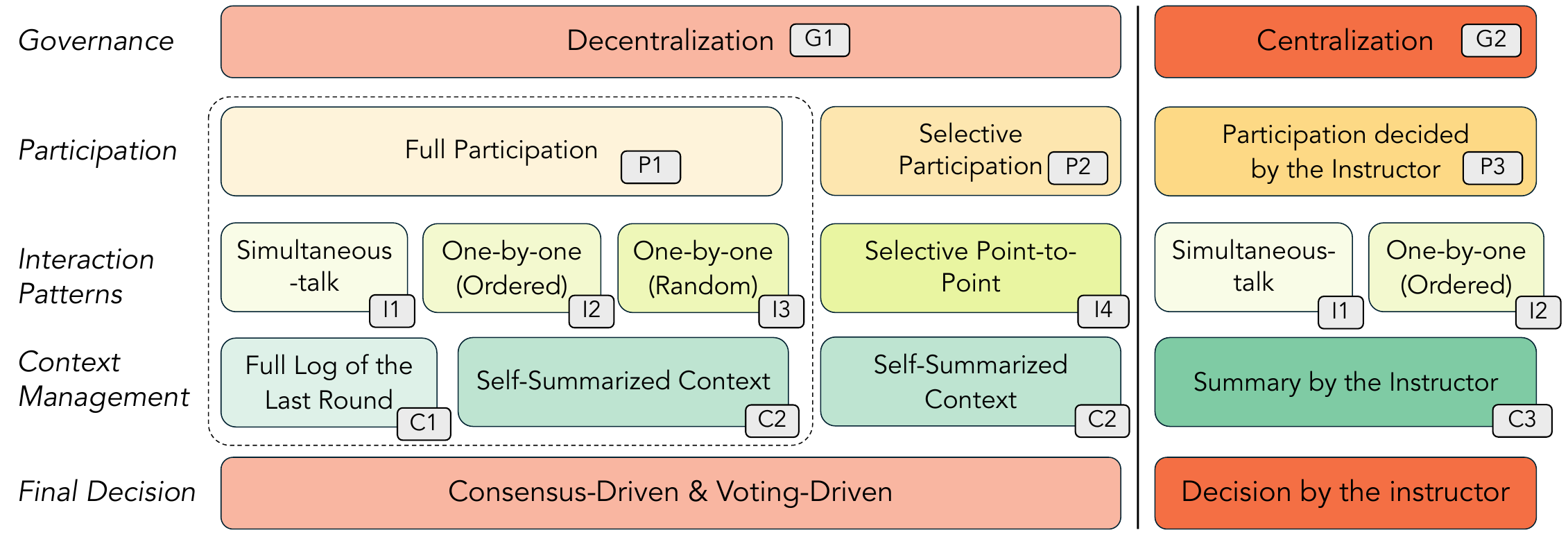}
    \caption{Combinations of collaboration strategies for multi-agent systems.}
    \label{fig:collaboration}
\end{figure*}

\section{Related Works}
\subsection{Multi-Agent Collaboration}
In real life, structured collectives of human individuals often demonstrate the ability to perform
tasks more efficiently and effectively through collaboration.
Drawing on this observation, AI researchers explore the potential of multi-agent systems~\cite{stone2000multiagent}
to improve task performance by enabling agents to work together in coordinated ways.
Recently, prior research has explored structures where agents interact sequentially to refine outputs.
Frameworks like \citet{chen2023agentverse} simulate sequential agent interactions to study
emergent behaviors, while \citet{chanchateval} and \citet{duimproving} employ debate-style
sequences to enhance reasoning and factuality. \citet{qian2025scaling} scales such systems by optimizing
turn-taking workflows for distributed tasks, and \citet{zhugegptswarm} dynamically adjusts
interaction graphs to balance sequential and parallel execution. These works highlight the benefits
of collaboration but often fix interaction patterns without analyzing how governance models
or participation rules impact efficiency.

\subsection{Role Specialization in Collaboration}
Role specialization underpins many multi-agent systems, where agents adopt domain-specific expertise.
\citet{kim2024mdagents} assigns medical roles for adaptive decision-making,
while \citet{zhang2024g} automates role-specific communication topologies via graph neural networks.
\citet{duimproving} and \citet{li2024more} advocate for role diversity in software teams, arguing
that ``more agents'' with distinct responsibilities improve outcomes. Similarly, \citet{su2024two}
uses targeted role subsets for scientific ideation. However, these works assume static role assignments
and overlook participation dynamics, such as when roles should contribute or how context should be
shared.

%% file: Text/3.collaboration.tex
\section{Collaboration Strategies for Multi-Agent System}

\subsection{Collaboration in Multi-Agent System}
Some previous research have employed a \textit{pipeline architecture}, where agents
sequentially process subtasks, passing outputs to downstream agents until the final goal is achieved~\cite{qian2025scaling, du2024multi}.
In contrast, our work investigates the collaboration among agents, each of which can
independently solve the task end-to-end, which mirrors real-world scenarios where domain experts—despite
individual competence—collaborate to integrate diverse perspectives, refine decisions, and mitigate blind spots.
Here, agents specialize in distinct \textit{aspects} of the task, and their collaboration aims
to synthesize these strengths rather than compensate for individual incapacity, further enabling dynamic
interaction strategies—governance, participation, interaction patterns, and context management—that
reflect human-like teamwork, as shown in Figure~\ref{fig:collaboration}.

\subsection{Governance: Decentralization or Centralization}
Governance of a multi-agent collaboration system, which is decentralization and centralization (marked as G1 and G2 in Figure~\ref{fig:collaboration}),
serves as the foundation for collaboration strategies, directly influencing participation, interaction patterns, and context management.

In decentralized governance, agents self-organize, autonomously decide when and how to participate,
interact, and manage context under specific rules on which all agents agree.
This fosters flexibility and scalability but may lead to coordination challenges, such as redundant contributions or a fragmented context.
Decisions in this framework often rely on collective approaches such as majority voting or consensus building.

As for centralized governance setting, an instructor agent oversees the discussion, dictating participation
(selecting which agents speak), interaction patterns (enforcing turn-taking or targeted
communication), context management (curating dialogue history for relevance) and controlled decision-making.
This ensures structured and efficient collaboration but risks bottlenecks if the instructor becomes
a single point of failure.

\subsection{Participation in Discussion Rounds}
Governance directly shapes participation strategies, determining which agents contribute and when.

\paragraph{Decentralization (G1)} Agents determine their own participation, leading to two strategies:

(i) \textbf{Full Participation} (G1-P1): All agents contribute in every round, ensuring diverse perspectives but potentially overwhelming the discussion with redundant or irrelevant inputs.

(ii) \textbf{Selective Participation} (G1-P2): Agents decide independently whether to speak and whom to address, based on their assessment of the needs of the discussion in the current discussion. For instance, an agent might choose to contribute only when its expertise is directly relevant or when it identifies a gap in the conversation. While this approach optimizes relevance, it risks overlooking critical inputs if agents misjudge the trajectory of the discussion.

\paragraph{Centralization (G2)} The instructor agent orchestrates the participation, explicitly deciding
which agents speak in each round and in what order (G2-P3).
This approach minimizes redundancy, but relies heavily on the ability of the instructor to
identify and sequence the most relevant agents.

\subsection{Interaction Patterns Among Agents}
Interaction patterns define how agents communicate with other agents during discussion rounds, with
governance and participation strategies shaping their design.
We identify four key patterns:

\paragraph{Simultaneous Talk (I1)}
All agents generate their responses simultaneously and independently within the same round, broadcasting their outputs to all peers.
Each agent has access to the complete discussion log from previous rounds.
This approach is suitable for scenarios involving the centralization setting (G2-P3) and full participation in the decentralization setting (G1-P1).
While it promotes diverse perspectives, it also carries the risk of conflicting or redundant output.

\paragraph{One-by-One (Ordered) (I2)}
Agents speak in a predefined sequence, either human-specified for decentralization (G1-P1) or instructor-enforced for centralization (G2-P3).
Each agent observes previous speech within the same round, enabling incremental refinement.

\paragraph{One-by-One (Random) (I3)}
Agents speak in a randomized sequence, also observing prior intra-round contributions.
This pattern, exclusive to decentralized governance with full participation (G1-P1), introduces stochasticity
to mitigate ordering biases, which prevents dominant agents from monopolizing discussions.

\paragraph{Selective Point-to-Point (I4)}
Agents autonomously decide whom to address, limiting communication to peers they deem relevant.
This pattern requires decentralized governance with selective participation (G1-P2), optimizing relevance
but risking fragmented context.

These relationships highlight how governance and participation constrain or enable interaction dynamics.
For instance, I4 is inherently decentralized, while I2 adapts to both governance models depending
on sequence control, as shown in Figure~\ref{fig:collaboration}.

\subsection{Context Management in Discussions}
As discussions progress, managing the growing dialogue history becomes critical to balance depth and computational efficiency.
Here, three strategies have been categorized including:

\paragraph{Full Log of the Last Round (C1)}
Agents retain the complete dialogue history from the most recent round, enabling comprehensive context
awareness, which has been also adopted in \citet{qian2025scaling}.
This method is typically adopted in decentralized systems with full participation (G1-P1), where all agents
contribute in every round and require full visibility into the prior discussion.
While this ensures rich context, it increases computational overhead and risks information overload.

\paragraph{Self-Summarized Context (C2)}
Each agent iteratively summarizes the discussion history, combining a condensed version of all prior
rounds with the full log of the last round.
This approach is suited for decentralized systems (G1), where agents independently manage context to
optimize relevance and efficiency.

\paragraph{Summary by the Instructor (C3)}
In centralized systems (G2-P3), the instructor agent summarizes the dialogue history for all participants.
This ensures consistency and relevance but introduces a single point of failure.

These strategies highlight the trade-offs between context depth and computational efficiency.
For instance, full log maximizes situational awareness but scales poorly, while self-summarized and
instructor-curated methods optimize efficiency at the cost of potential information loss.

\subsection{Final Decision Mechanisms}
The process of terminating multi-agent discussion and making a final decision is tightly coupled
with governance models, therefore formalized as:

\paragraph{Decentralization}
Decentralized systems rely on consensus or majority voting.
Agents autonomously detect agreement based on the generated prediction and terminate discussions once consensus emerges.
And if consensus is not reached within a predefined maximum round limit, agents trigger majority voting to force a decision.

\paragraph{Centralization}
In centralized systems, the instructor agent determines when to  finalize the discussion.
The instructor evaluates the discussion progress against predefined criteria and either continues the discussion
if critical disagreements persist, or terminates the discussion and selects the final decision.

The complete set of possible permutations of the collaboration settings can be found in Appendix~\ref{collaboration-setting}.

%% file: Text/4.experiment.tex
\section{Experiments}
\subsection{Experimental Design}
\paragraph{Context-Based Collaboration}
In our experimental framework, we seek to evaluate whether LLM-based multi-agents can accurately
tackle problems by leveraging specific contextual information from agents.
Prior work often defines agent roles implicitly via prompts (e.g.\ ``You are a radiologist'')~\cite{kim2024mdagents, qian2025scaling},
which conflates role assignment with specialization but fails to measure the proficiency of agents
in leveraging domain knowledge.
To address this, we explicitly equip agents with distinct prior knowledge, which is task-specific
context segments that constrain their inputs and outputs, ensuring collaboration arises from complementary
expertise rather than role labels (see more details in Appendix~\ref{context-aware}).
\paragraph{Evaluation Metrics}
To comprehensively assess the performance of LLM-based multi-agent systems, we examine both task accuracy and computational cost efficiency.
(1)~Accuracy, which is measured as the rate at which the multi-agent system arrives at the correct classification.
(2)~Token Count, including the input and output token count, captures the mean total
number of tokens processed as input and output by the agents during all discussion rounds as an indicator
of the volume of contextual information handled by the system.
(3)~Discussion Rounds, which is the number of communication rounds required for the agents to converge
on a final decision, indicating how quickly the multi-agent system reaches a consensus.
By jointly considering both accuracy and these computational cost metrics, we aim to provide an
evaluation of our multi-agent framework, balancing the trade-offs between decision quality and resource efficiency.

As for backbone models, we conduct all experiments on ChatGPT-4o\footnote{ChatGPT-4o-0806 version}.

\subsubsection{Context-Dependent Task Selection}
We adopt two tasks requiring agents to ground decisions strictly in provided contexts, minimizing
reliance on the internal knowledge of LLMs.
The first task, termed ``Distributed Evidence Integration (DEI)'', challenges agents to collaboratively
combine fragmented pieces of evidence—each drawn from distinct context segments—to arrive at a unified decision.
In contrast, the second task, defined as ``Structured Evidence Synthesis (SES)'', requires agents to
critically assess and synthesize pre-labeled pieces of evidence, with each agent assigned a single
element, to verify factual claims.
Together, these tasks emphasize context-based decision-making while fostering robust multi-agent collaboration:

\paragraph{Distributed Evidence Integration (DEI)}
\label{pddp-intro}
For the DEI scenario, we utilize the MIMIC-III dataset~\cite{johnson2016mimic}, a comprehensive,
publicly available database of de-identified health-related information including clinical records, vital signs, medications,
and diagnoses, primarily from intensive care unit (ICU) patients.
The discharge summary notes, in particular, encapsulate key patient details, such as the brief hospital course.

In this task, a cohort of agents is each assigned a distinct clinical context segment (i.e., brief
hospital course, major surgery or invasive procedure, pertinent results, or social history) and
is tasked with predicting the patient’s discharge disposition among four possible outcomes: ``expired'',
``extended care'', ``home with service'', and ``home'' as task of Patient Discharge Disposition Prediction (PDDP)
(see Appendix~\ref{label_pddp} for more details).
Notably, no explicit evidence label is provided to any individual agent.
Instead, the PDDP task requires that all agents collaboratively integrate their partial information
to arrive at a final consensus.

\paragraph{Structured Evidence Synthesis (SES)}
\label{ebfc-intro}
For the SES scenario, an evidence-based fact-checking tasks (EBFC) leverages the AMBIFC
dataset~\cite{glockner2024ambifc} to evaluate the fact-checking capabilities of
multi-agent systems by synthesizing evidence exclusively from the provided contextual data.
In this dataset, each claim is accompanied by numerous evidence sentences, yet only a small subset is directly relevant to the claim.
Furthermore, each agent is assigned a single piece of evidence (an example is provided in Appendix~\ref{EBFC}).
This setup compels the agents to engage in collaborative negotiation: those who receive evidence directly pertinent to the
claim must persuade their peers—who may have been allocated less relevant evidence—to converge on an accurate, factually sound assessment.
Thus, while the PDDP task emphasizes collective deliberation in the absence of explicit evidence,
the EBFC task challenges agents to build consensus by leveraging and disseminating critical evidence
held by only a subset of agents.

\subsection{Performance on Individual Agent}
Before conducting experiments on multi-agent collaboration, we first evaluate the performance of individual agents.

For the DEI scenario, as Table~\ref{tab:pddp-single} the agent equipped with the brief hospital course context (\(Agent_{\text{BHC}}\))
achieves the best performance, as it has access to a comprehensive overview of the patient's trajectory.
In contrast, other agents often lack essential details, leading to suboptimal performance.
Moreover, the \(Agent_{\text{all}}\)—which aggregates information from multiple sources—performed slightly
worse than \(Agent_{\text{BHC}}\) due to the interference of misleading details from other agents.
This misleading effect becomes even more pronounced when the final decision is determined through majority voting.
Consequently, for the DEI scenario, it is anticipated that collaboration will be most effective when
other agents contribute complementary information to \(Agent_{\text{BHC}}\) for a holistic decision.

\begin{table}[htbp]
\centering
\resizebox{0.90\columnwidth}{!}{
\begin{tabular}{c|c|c|c|c}
\hline
\textbf{Methods} & \textbf{Acc}$\uparrow$ & \textbf{$\#$I}$\downarrow$ & \textbf{$\#$O}$\downarrow$ & \textbf{Round}$\downarrow$ \\
\hline
$Agent_{\text{BHC}}$ & 60.8  & 541 & 109 & 1 \\
$Agent_{\text{MSIP}}$ & 38.7  & 170 & 91 & 1 \\
$Agent_{\text{PR}}$ & 33.7  & 492 & 119 & 1 \\
$Agent_{\text{DM}}$ & 39.2  & 488 & 125 & 1 \\
$Agent_{\text{SH}}$ & 41.2  & 182 & 88 & 1 \\ \hline
$Agent_{\text{all}}$ & 57.8  & 1,281 & 129 & 1 \\
MV & 47.2  & 1,873 & 661 & 5 \\
\hline
\end{tabular}
}
\caption{Performance of individual agent on the PDDP task. \#I means input token count. \#O means output token count.
BHC: Brief Hospital Course. MSIP: Major Surgical or Invasive Procedure. PR: Pertinent Results.
DM: Discharge Medications. SH: Social History. $Agent_{\text{all}}$: inference by an individual agent with all information concatenated.
MV: major voting of all agents.}
\label{tab:pddp-single}
\end{table}

For the SES scenario, experimental results in Table~\ref{tab:ebfc-single} reveal that the agent receiving relevant evidence (\(Agent_{\text{consistent}}\)) attains an accuracy of 88.7\%.
Notably, the 88.7\% accuracy represents the \textit{theoretical upper bound for the multi-agent system's performance}.
Interestingly, \(Agent_{\text{all}}\) achieves even higher accuracy than \(Agent_{\text{consistent}}\), which can be attributed to the randomness introduced by aggregating inputs from all agents.
In contrast, the agent provided with irrelevant evidence (\(Agent_{\text{inconsistent}}\)) performs no better than a random guess.
In this scenario, majority voting is suitable, as the majority of agents possess inconsequential information.
Therefore, for the SES scenario, it is expected that \(Agent_{\text{consistent}}\) should lead the discussion,
effectively persuading other agents to converge on the correct assessment.

\begin{table}[htbp]
\centering
\resizebox{0.90\columnwidth}{!}{
\begin{tabular}{c|c|c|c|c}
\hline
\textbf{Methods} & \textbf{Acc}$\uparrow$ & \textbf{$\#$I}$\downarrow$ & \textbf{$\#$O}$\downarrow$ & \textbf{Round}$\downarrow$ \\
\hline
$Agent_{\text{consistent}}$ & 88.7  & 177 & 48 & 1 \\
$Agent_{\text{inconsistent}}$ & 19.2  & 172 & 56 & 1 \\ \hline
$Agent_{\text{all}}$ & 90.5  & 283 & 55 & 1 \\
MV & N/A & N/A & N/A & N/A \\
\hline
\end{tabular}
}
\caption{Performance of individual agent on the EBFC task.  \#I means input token count. \#O means output token count.
``consistent'': individual agent infers only with single evidence with the same label as that of the claim.
``inconsistent'': individual agent infers only with single evidence with different label from that of the claim.
$Agent_{\text{all}}$: inference by an individual agent with all evidence concatenated.
MV: major voting of all agents. N/A: not applicable.}
\label{tab:ebfc-single}
\end{table}

\begin{table}[htb]
\centering
\resizebox{0.90\columnwidth}{!}{
\begin{tabular}{c|c|c|c|c}
\hline
\textbf{Methods} & \textbf{Acc}$\uparrow$ & \textbf{$\#$I}$\downarrow$ & \textbf{$\#$O}$\downarrow$ & \textbf{Round}$\downarrow$ \\
\hline
$Agent_{\text{all}}$              & 57.8    & 1,281       & 129         & 1.00      \\
MV & 47.2    & 1,873       & 661         & 5.00     \\
\hline
G1-P1-I1-C1  & 50.7 & 25,663   & 2,184     & 3.28   \\
G1-P1-I2-C1  & 57.8 & \underline{6,470}    & \underline{854}      & \underline{1.30}   \\
G1-P1-I3-C1  & 45.2 & 9,531    & 1,127     & 1.65   \\
G1-P1-I1-C2  & 46.2 & 52,400   & 15,568    & 4.43   \\
G1-P1-I2-C2  & \textbf{59.8} & 15,057   & 3,046     & 1.56   \\
G1-P1-I3-C2  & 46.7 & 19,673   & 4,100     & 1.81   \\
G1-P2-I4-C2  & 50.8 & 348,035  & 58,795    & 9.91   \\
G2-P3-I1-C3  & 46.2 & 13,119   & 2,412     & 2.04   \\
G2-P3-I2-C3  & \underline{58.8}  & \textbf{4,867}  & \textbf{841}    & \textbf{1.03}  \\
\hline
\end{tabular}
}
\caption{Performance of multi-agent collaboration on the PDDP task with different collaboration strategies.
\#I means input token count. \#O means output token count. $Agent_{\text{all}}$: inference by an individual
agent with all information concatenated. MV: major voting of all agents. The best and the second-best results
are in \textbf{bold} and \underline{underlined}.}
\label{pddp-multi-agent}
\end{table}

\begin{table}[htb]
\centering
\resizebox{0.90\columnwidth}{!}{
\begin{tabular}{c|c|c|c|c}
\hline
\textbf{Methods} & \textbf{Acc}$\uparrow$ & \textbf{$\#$I}$\downarrow$ & \textbf{$\#$O}$\downarrow$ & \textbf{Round}$\downarrow$ \\
\hline
$Agent_{\text{consistent}}$ & 88.7    & 177        & 48          & 1.00   \\
\hline
G1-P1-I1-C1   & 49.3  & 28,099    & 1,990     & 6.28 \\
G1-P1-I2-C1   & 70.4  & 13,361    & 1,155      & 3.35 \\
G1-P1-I3-C1  & 68.8  & 15,368    & 1,284      & 3.99 \\
G1-P1-I1-C2   & 81.4  & 20,074    & 6,780      & 3.46 \\
G1-P1-I2-C2   & 84.4  & 14,600    & 3,073      & 2.00 \\
G1-P1-I3-C2   & 77.9  & 13,125    & 2,901      & 2.14 \\
G1-P2-I4-C2   & \underline{86.4}  & 30,085    & 6,125      & 2.76 \\
G2-P3-I1-C3  & \textbf{86.9}  & \textbf{2,111}     & \underline{490}       & \underline{1.16} \\
G2-P3-I2-C3  & 85.4  & \underline{2,859}     & \textbf{452}       & \textbf{1.10} \\ \hline
\end{tabular}
}
\caption{Performance of multi-agent collaboration on the EBFC task with different collaboration strategies.
\#I means input token count. \#O means output token count.
$Agent_{\text{consistent}}$ is the theoretical upper bound of the multi-agent systems.
The best and the second-best results are in \textbf{bold} and \underline{underlined}.}
\label{ebfc-multi-agent}
\end{table}

\begin{figure}[tb]
    \centering
    \includegraphics[width=0.78\columnwidth]{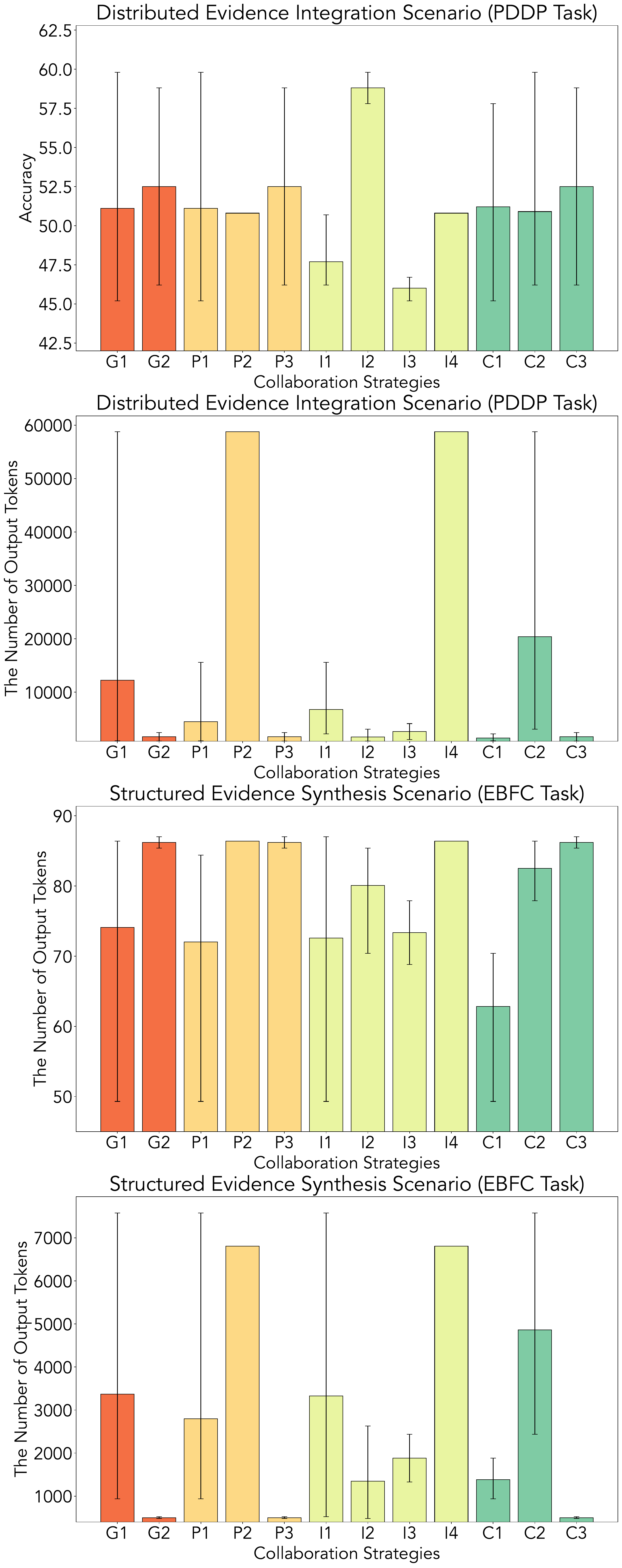}
    \caption{Performance of multi-agent systems on the PDDP and EBFC tasks considering individual strategy dimensions. Error bars here mark the maximum and minimum values.}
    \label{fig:pddp-ebfc}
\end{figure}

\subsection{Performance on Multi-Agent Systems}
\label{sec:para-tar}

Tables~\ref{pddp-multi-agent} and~\ref{ebfc-multi-agent} illustrate the performance of multi-agent
collaboration across various strategies for the DEI and SES scenarios.
The results highlight that different configurations of collaboration dimensions significantly
influence the performance, leading to an accuracy gap of up to 37.6\% (ranging from 49.3\% to 86.9\%).
Despite some strategies achieving similar performance in terms of accuracy, there is a substantial variation in token cost.
For instance, in the EBFC task (SES), the strategy ``G1-P2-I4-C2'' achieves an accuracy of 86.4\%,
comparable to that of ``G2-P3-I1-C3'', but costs 11.5 times more output tokens, demonstrating a significant difference in computational cost.

For a more in-depth analysis, Figure~\ref{fig:pddp-ebfc} presents the average accuracy and output
token count for a single dimension of the collaboration settings, covering governance, participation,
interaction patterns, and context management in discussion.
The maximum and minimum values are also annotated using error bars, providing further insight into the variability across strategies.
These visualizations reveal both similarities and differences in collaboration strategies between the DEI and SES scenarios.

\paragraph{DEI Scenario}
In the DEI scenario, the final prediction relies on distributed evidence among agents, requiring them to consolidate information.
Consequently, multi-agent systems are expected to outperform individual agents.
However, as shown in Table~\ref{pddp-multi-agent}, only two out of nine multi-agent systems achieve this goal,
further emphasizing the importance of nuanced collaboration strategies.

\paragraph{SES Scenario}
For the SES scenario, only a small subset of agents hold relevant evidence, which may lead to misguidance by agents without useful evidence.
As a result, the accuracy of $Agent_{\text{consistent}}$ represents the theoretical upper bound for multi-agent systems.
Table~\ref{ebfc-multi-agent} illustrates that the strategy ``G2-P3-I1-C3'' achieves an accuracy most
closely aligned with that of $Agent_{\text{consistent}}$.

\paragraph{Governance}
In the PDDP task, Figure~\ref{fig:pddp-ebfc} demonstrates that the governance dimension, whether
decentralized (G1) or centralized (G2), does not significantly affect the accuracy (mean, max, or min).
However, G1 tends to have much higher mean and maximum output token counts compared to G2, indicating
that decentralized governance leads to higher token cost in the DEI setting.
The lack of significant accuracy differences between G1 and G2 suggests that DEI does not inherently favor hierarchical control.
However, G1’s higher token costs reflect the merits of instructor-mediated coordination, which enables
agents to organize around distributed clinical data (e.g.,\ lab results, surgical history), reducing redundancy.

Similarly, for the EBFC task, G1 and G2 show similar maximum accuracy, but the accuracy fluctuation
in G2 is notably smaller because the instructor is more readily persuaded by the relevant evidence.
Token costs also follow a similar pattern to the PDDP task, with G2 demonstrating greater efficiency in terms of output token usage.

\paragraph{Participation}
For the PDDP task, Table~\ref{pddp-multi-agent} reveals that P1 (Full Participation) and P3 (Participation
decided by the instructor) exhibit higher accuracy ceilings than P2 (Selective Participation), with a gap of up to 9.0\%
because discharge disposition prediction requires synthesizing \textit{all} contextual facets.
P2 may risk omitting critical inputs, such as vital signs due to misjudged relevance, while P3
mitigates this risk by leveraging global attention to key indicators.
Nonetheless, the performance of these strategies fluctuates significantly depending on other factors in the collaboration setup.

For the EBFC task, P2 and P3 not only perform better but also with less fluctuation, and P2’s superior
performance reflects the task s demand for targeted expertise.
In terms of token cost, P2 consistently requires more tokens compared to P1 and P3, with P1 costing
the highest maximum output token count in the EBFC task.
P3, however, consistently costs the least token cost in both tasks, which further highlights the
instructor’s role in suppressing redundant contributions.

\paragraph{Interaction}
In the PDDP task, I2 (Ordered One-by-one) outperforms all other interaction settings, delivering superior accuracy and
output token efficiency, suggesting that it may be the optimal interaction strategy.
I1 probably introduces conflicting evidence at the same time which leads to noise conclusion, while
I4 may fragment essential contexts from other agents unconsciously.

For the EBFC task, differences in mean accuracy across interaction settings are minimal, suggesting
SES scenario tolerates flexible interaction strategies.
However, in terms of output token count, I2 and I3 (Random One-by-one) perform similarly, both showing
substantial improvements over I1 (Simultaneous-talk) and I4 (Selective Point-to-Point), stemming
from organized evidence reconciliation and possible force redundant backtracking in the dialog rounds.

\paragraph{Context}
For the PDDP task, context management strategies do not show a significant difference in terms of
accuracy, though C1 (Full Log of the Last Round) and C3 (Summary by the Instructor) outperform C2
(Self-Summarized Context) with token count, indicating that DEI benefits both from comprehensive
context (C1) and distilled insights (C3).
However, self-summarized context lags due to inconsistent truncation of critical details.

In the EBFC task, C2 and C3 achieve over 16.0\% higher accuracy than C1, and C3 shows more stable
performance, arising from the instructor’s ability to highlight salient evidence.
However, C2, which overemphasizes an agent’s preferred evidence type, costs much higher token costs
compared to both C1 and C3, making C3 the optimal setting for the SES scenario.\

\paragraph{Token-Accuracy-Ratio (TAR)}

To comprehensively evaluate the performance of these collaboration strategies, we introduce the Token-Accuracy Ratio (TAR),
which accounts for the computational efficiency in terms of accuracy along with both input and output tokens.
The formula for TAR is:

\begin{equation}
    \text{TAR} = \frac{\text{Accuracy}}{\alpha \cdot \#\text{I} + \beta \cdot \#\text{O}}
\end{equation}

\noindent where \(\alpha\) and \(\beta\) are coefficients for the relative computational cost of
input and output tokens, respectively.
Based on the pricing of ChatGPT 4o, where the cost for output tokens is four times that of input tokens\footnote{https://openai.com/api/pricing/},
we set \(\alpha = 1\) and \(\beta = 4\).
The detailed results are provided in Appendix~\ref{subsec:tar}.

From Table~\ref{pddp-tar} and~\ref{ebfc-tar}, the strategy ``G2-P3-C3'' achieves the optimal TAR across
both scenarios, indicating that multi-agent collaboration with an instructor overseeing participation, context,
and the final decision tends to yield optimal performance under both the DEI and SES settings.

\subsection{Summary}
Our analysis across various strategies reveals significant trade-offs between accuracy and computational cost.
Notably, the introduction of the Token-Accuracy Ratio (TAR) highlights that configurations such as ``G2-P3-C3''
offer an optimal balance, underscoring the importance of nuanced design choices in enhancing both decision quality and efficiency in multi-agent systems.

%% file: Text/6.conclusion_ethics.tex
\section{Conclusion}
In this study, we systematically investigate the understudied the fine-grained mechanics of collaboration
in multi-agent systems, focusing on governance, participation, interaction patterns, and context management.
Through experiments on two tasks under two scenarios correspondingly, we demonstrate that centralized governance,
guided by an instructor agent, consistently balances accuracy and computational efficiency.
By shifting focus from structural novelty to strategic collaboration, this work provides a foundation
for designing efficient, scalable, and context-aware multi-agent systems.

%% file: Text/7.Limitations.tex
\section*{Limitations}
This study has several limitations.
First, the performance of the multi-agent system depends on the quality and completeness of the contextual data provided to each agent.
Incomplete or ambiguous data can affect decision-making accuracy.
Second, scalability issues may arise as the number of agents increases, with potential challenges in coordination and computational efficiency.
Additionally, our framework assumes agents can independently interpret context, which may be influenced by biases or lack of domain-specific knowledge.
The generalization of our approach across different domains remains uncertain, as the tasks used are specific to certain contexts.
Moreover, our system does not integrate external knowledge, which could limit performance in dynamic or evolving scenarios.
Lastly, the interpretability of multi-agent decisions remains a challenge, as the rationale behind agent interactions can be difficult to trace.
Despite these limitations, the study lays a foundation for further strategy design of multi-agent collaboration in decision-making tasks.

\section*{Ethical Considerations}
This study uses the MIMIC-III dataset, which contains de-identified ICU patient data.
We ensure all data usage complies with ethical guidelines and privacy standards, as no personally identifiable information is included.
Additionally, while our multi-agent systems are designed to assist in decision-making, they should
not replace human judgment in critical healthcare contexts.
We emphasize the importance of transparency and accountability in the deployment of AI systems in healthcare.

%% file: Text/x.Appendix.tex
\begin{figure*}[htb]
    \centering
    \includegraphics[width=0.99\textwidth]{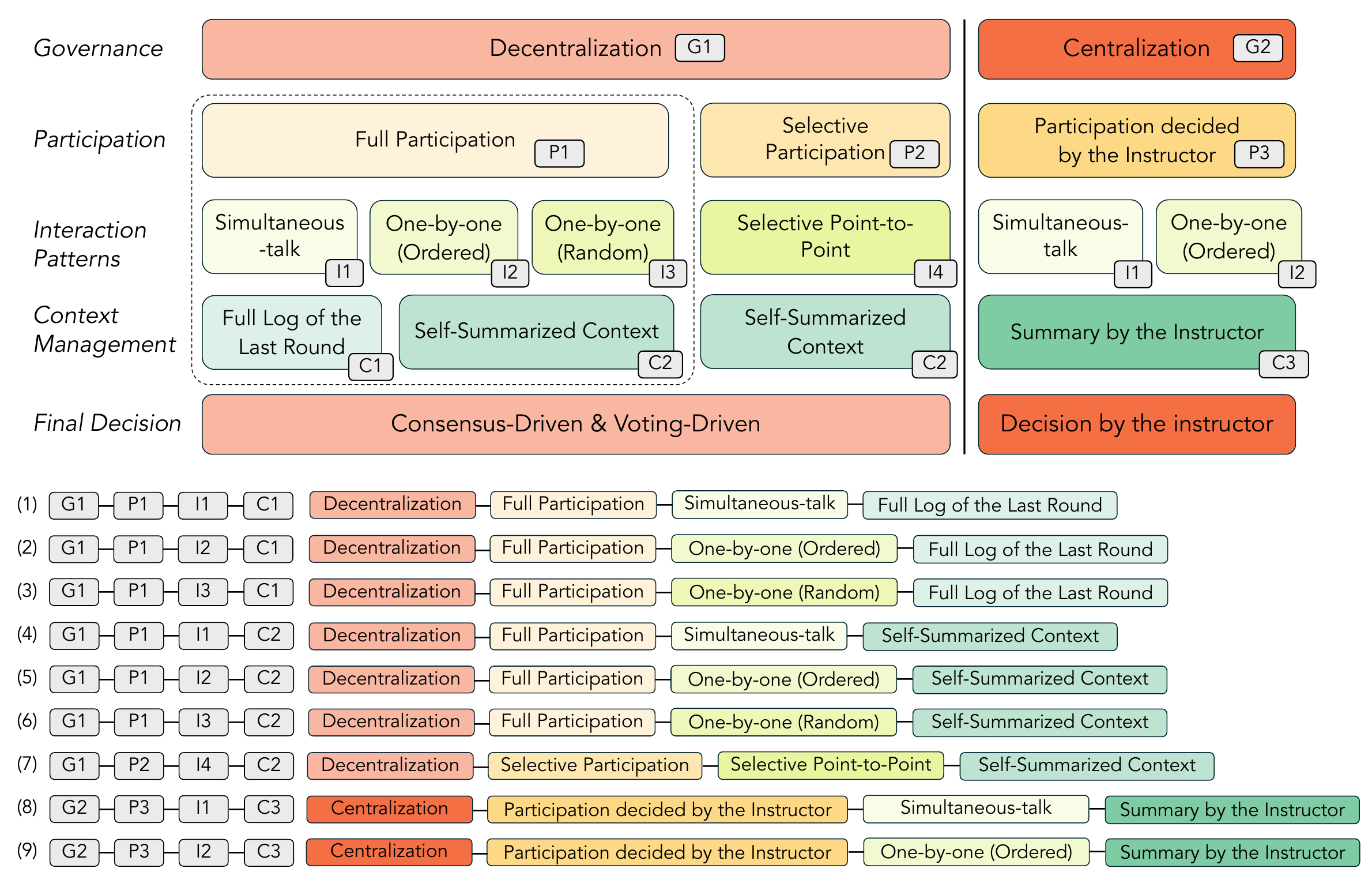}
    \caption{The complete set of possible permutations of the collaboration settings.}
    \label{all_permutation}
\end{figure*}

\appendix

\section{Collaboration Strategies}
\label{collaboration-setting}
The complete set of possible permutations of the collaboration settings is in Figure~\ref{all_permutation}.

\begin{figure*}[thb]
    \centering
    \includegraphics[width=0.99\textwidth]{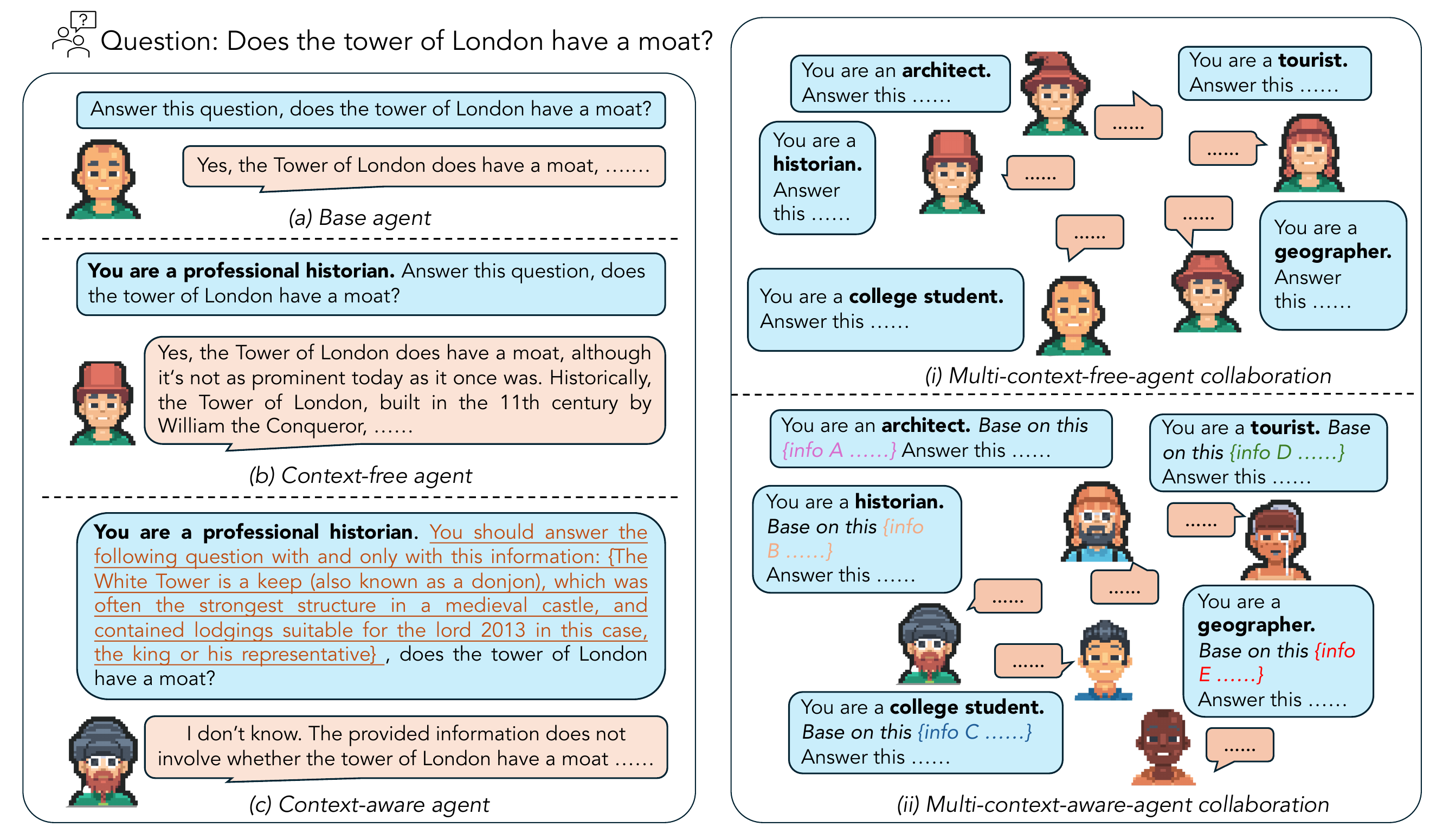}
    \caption{Illustration for multi-context-based agent collaboration.}
    \label{intro_case}
\end{figure*}
\section{Context-Based Collaboration}
\label{context-aware}

As Figure~\ref{intro_case}, a context-free agent generates responses based on general knowledge
without restriction, allowing it to incorporate external background information.
In contrast, a context-based agent strictly adheres to the given context, responding only with
the provided information, even if it lacks sufficient details to answer the query.
As shown in Figure~\ref{intro_case}, when asked whether the Tower of London has a moat, the context-free
agent leverages historical knowledge to provide an answer, whereas the context-based agent, constrained
by the given excerpt, acknowledges its inability to determine the answer.
This behavior closely mirrors real-life human reasoning, where individuals rely on the information
at hand rather than external knowledge.
More importantly, in the analysis of multi-agent collaboration strategies, context-based agents
enable a clearer investigation of how agents interact and share information without interference
from the internal knowledge embedded in LLMs as shown in Figure~\ref{intro_case}.
By eliminating this confounding factor, context-based agents provide a more controlled setting for
studying knowledge exchange, reasoning dynamics, and the emergence of cooperative problem-solving
in multi-agent systems.

\section{Details for Patient Discharge Disposition Prediction Task}
\label{label_pddp}
\subsection{Label Definition}
In the context of the discharge disposition of patients in the MIMIC-III dataset, the following terms represent various settings in which a patient may be discharged from the hospital, reflecting the type of care they will receive after their discharge:
\begin{enumerate}
\item Expired: This refers to patients who died during their hospital stay. The discharge disposition is marked as ``expired'' when the patient is no longer alive at the time of discharge.

\item Extended Care: This refers to patients who are discharged to a facility that provides longer-term care than the acute hospital setting, but not as intensive as inpatient care. These facilities often include skilled nursing facilities, rehabilitation centers, or similar establishments that provide continued medical care, physical therapy, or recovery support.

\item Home with Service: This indicates that the patient is discharged home but will continue to receive some form of medical care or assistance. This could involve home health services such as nursing care, physical therapy, or other medical support delivered in the patient's home.

\item Home: This refers to patients who are discharged directly to their home, without the need for continued medical care or services. These patients typically no longer require hospitalization or any ongoing treatment and are considered well enough to return to normal activities.

\end{enumerate}
These categories are essential in capturing the outcomes and planning for a patient's post-discharge care, as they can significantly influence the patient's recovery trajectory and healthcare planning.

\subsection{Task Example}
In Section~\ref{pddp-intro}, distributed evidence integration is referred to the process by which multiple agents,
each possessing only a fragment of the overall evidence, collaboratively combine their partial information
to reach a comprehensive and accurate decision.
In the example shown in Table~\ref{pddp-example}, the complete set of evidence regarding the patient’s
case is distributed across several distinct context segments—such as the Brief Hospital Course,
Major Surgical or Invasive Procedure, Pertinent Results, Discharge Medications, and Social History.
Each segment represents information held by a different agent.
No single agent has access to all the details necessary to determine the correct Discharge Disposition (in this case, ``Extended Care'').

Through the process of distributed evidence integration, the agents must share and synthesize their
individual pieces of evidence, engaging in discussion and negotiation to resolve discrepancies and fill gaps in the information.
This collaborative integration ensures that the final decision is informed by all available evidence,
thereby leveraging the strengths of each agent’s specialized knowledge.

\begin{table*}[htb]
\centering
\begin{tabularx}{\textwidth}{p{0.1\textwidth}|p{0.85\textwidth}}
\hline
Brief Hospital Course & Ms.  was admitted to  on  for treatment of right brain tumor. She was plegic on the left side and was taken to the OR on  by Dr. . Post-op CT scan was stable. Decadron 4mg every 6 hours was continued, and on  she was cleared to transfer to the floor. On  her MRI showed some residual tumor but decreased midline shift. She regained some strength in the LLE. The sterroid was subsequently tapered to 2 mg .  PT and OT were consulted and recommended rehab. Radiation Oncology was consulted and the patient will follow-up for radiation treatment after discharge. Neuro-oncology was made aware of Ms.  and the patient has a Brain  Clinic appointment with them to discuss chemotherapy after the final pathology is back. The patient was discharged to rehab on . \\  \hline
Major Surgical or Invasive Procedure & Right craniotomy for tumor resection \\  \hline
Pertinent Results & MRI brain : There is a large heterogeneously rim-enhancing mass in the right frontal lobe measuring 2.4 x 2.7 cm with enhancement also extending to the right frontal  subependymally and to the corpus callosum. An additional focus of enhancing abnormality is seen in the right temporal lobe.There is edema surrounding the right frontal lobe lesion with midline shift.  MRI brain : Patient is status post resection of right frontal heterogeneously enhancing mass. There are blood products in the operative bed, which limit evaluation for residual neoplasm. However, there does appear to be residual enhancing abnormality in the right frontal lobe and along the inferior margin of the operative cavity extending into the corpus callosum and the caudate head.  There is also nodular enhancement along the anterior and posterior margins of the operative cavity superiorly. There is a focus of restricted diffusion in the right frontal lobe along the inferior lateral margin of the cavity. This may represent cytotoxic edema from surgery.  There is a small right hemispheric extra-axial postoperative collection. Right parafalcine extra-axial collection is also noted.There is slight improvement in the midline shift to the left. \\  \hline
Discharge Medications & 1. Acetaminophen 325 mg Tablet Sig: 1-2 Tablets PO Q6H (every 6 hours) as needed for pain/t/HA. 2. Docusate Sodium 100 mg Capsule Sig: One (1) Capsule PO BID (2 times a day). 3. Aripiprazole 10 mg Tablet Sig: Two (2) Tablet PO DAILY (Daily). 4. Oxcarbazepine 600 mg Tablet Sig: One (1) Tablet PO HS (at bedtime). 5. Clonazepam 1 mg Tablet Sig: One (1) Tablet PO BID (2 times a day). 6. Atorvastatin 20 mg Tablet Sig: One (1) Tablet PO DAILY (Daily). 7. Insulin Lispro 100 unit/mL Solution Sig: One (1) Subcutaneous ASDIR (AS DIRECTED). 8. Nicotine 14 mg/24 hr Patch 24 hr Sig: One (1) Patch 24 hr Transdermal DAILY (Daily). 9. Sertraline 50 mg Tablet Sig: Four (4) Tablet PO QHS (once a day (at bedtime)). 10. Quetiapine 25 mg Tablet Sig: Two (2) Tablet PO QAM (once a day (in the morning)). 11. Quetiapine 300 mg Tablet Sustained Release 24 hr Sig: Two (2) Tablet Sustained Release 24 hr PO QHS (once a day (at bedtime)). 12. Oxycodone-Acetaminophen 5-325 mg Tablet Sig: 1-2 Tablets PO Q4H (every 4 hours) as needed for Pain. 13. Levetiracetam 500 mg Tablet Sig: Two (2) Tablet PO BID (2 times a day). 14. Heparin (Porcine) 5,000 unit/mL Solution Sig: One (1) Injection TID (3 times a day). 15. Dexamethasone 2 mg Tablet Sig: One (1) Tablet PO Q8 hours () for 5 doses. 16. Dexamethasone 2 mg Tablet Sig: One (1) Tablet PO Q12 hours (): Please start after 2 Q8 hour dose is complete. \\  \hline
Social History & social ETOH, 15 cigarettes per day. works as dishwasher and typer, lives alone \\  \hline
Discharge Disposition & Extended Care \\ \hline
\end{tabularx}
\caption{An example of PDDP task. }
\label{pddp-example}
\end{table*}

\section{Details for Evidence-Based Fact-Checking Task}
\label{EBFC}
In Section~\ref{ebfc-intro}, structured evidence synthesis is referred to the systematic process
by which agents analyze and combine pre-labeled pieces of evidence to determine the veracity of a given claim.
In the example shown in Table~\ref{ebfc-example}, the claim (``Season 5 the last season was of Ray Donovan'')
is accompanied by multiple evidence sentences.
Each evidence sentence is annotated with a label—such as ``Refuting'' or ``Neutral''—that indicates
its relevance or relation to the claim.

In this task, each agent is provided with a single piece of such structured evidence.
The challenge lies in the agents' ability to engage in dialogue and collectively synthesize the relevant
information from these distributed, labeled pieces of evidence.
Agents holding evidence that directly pertains to the claim (e.g.\ the sentence indicating that the
series was canceled after seven seasons) must persuade other agents—who might have received less
directly relevant or neutral evidence—to recognize the overall factual context.
Through this collaborative synthesis, the agents work together to arrive at a well-supported, final decision regarding the claim.

Thus, structured evidence synthesis emphasizes:
(1) The use of clearly labeled, structured evidence.
(2) The necessity for agents to extract, assess, and combine this evidence.
(3) The collaborative negotiation required to integrate disparate information into a coherent conclusion.
\begin{table*}[htb]
\centering
\begin{tabularx}{\textwidth}{p{0.12\textwidth}|p{0.7\textwidth}|p{0.12\textwidth}}
\hline
\textbf{Type} & \textbf{Sentence} & \textbf{Label} \\ \hline
Claim & Season 5 the last season was of ray donovan. & Refuting \\ \hline
Evidence 1 & On February 4, 2020, Showtime cancelled the series after seven seasons. & Refuting \\ \hline
Evidence 2 & The twelve-episode first season premiered on June 30, 2013. & Neutral\\ \hline
Evidence 3 & The pilot episode broke viewership records, becoming the biggest premiere of all time on Showtime.  & Neutral\\ \hline
Evidence 4 & The show was cancelled without any advance warning, leaving fans and showrunner, David Hollander, in shock. & Neutral\\ \hline
Evidence 5 & A week later, Liev Schreiber commented on his Instagram that due to fans' support and activity in media, there will be more Ray Donovan. & Neutral\\ \hline
Evidence 6 & The drama is set primarily in Los Angeles, California (during seasons 1–5) and primarily in New York City, New York (during seasons 6–7). & Neutral\\ \hline
\end{tabularx}
\caption{An example of EBFC task. }
\label{ebfc-example}
\end{table*}

\section{Details for Token-Accuracy-Ratio}
\label{subsec:tar}
To holistically evaluate the performance of multi-agent collaboration strategies, we introduce
the Token-Accuracy Ratio (TAR), a metric that balances task accuracy against computational cost.
The TAR is defined as mentioned in Section~\ref{sec:para-tar}:
\begin{equation}
    \text{TAR} = \frac{\text{Accuracy}}{\alpha \cdot \#\text{I} + \beta \cdot \#\text{O}}
\end{equation}

To facilitate cross-task comparisons, we also define the Normalized TAR (NTAR), which scales the
TAR values relative to the maximum TAR observed for each task.
This normalization ensures that the results are comparable across tasks with different accuracy and token cost ranges.
The Normalized TAR is calculated as:

\[
\text{Normalized TAR} = \frac{\text{TAR}}{\text{Max TAR for the task}}
\]

\subsection{PDDP Task (Distributed Evidence Integration)}
In the PDDP task, agents collaborate to predict patient discharge outcomes by integrating fragmented clinical data.
The results highlight the trade-offs between accuracy and computational cost:

\paragraph{Optimal Strategy} The strategy ``G2-P3-I2-C3'' (centralized governance, instructor-led participation,
ordered interaction, and instructor-curated context summarization) achieves the highest Normalized TAR of 1.0,
with an accuracy of 58.8\% and the lowest token counts ($\#$Input Token 4,867 and $\#$Output Token 841).
This configuration demonstrates the efficiency of centralized control in reducing redundancy and optimizing resource usage.

\paragraph{Decentralized Strategies} Decentralized configurations, such as ``G1-P1-I2-C1'' and ``G1-P1-I2-C2''
achieve competitive accuracy (57.8\% and 59.8\%, respectively) but cost significantly higher token costs.
For example, ``G1-P1-I2-C2'' achieves the highest accuracy (59.8\%) but requires 15,057 input tokens and 3,046 output tokens,
resulting in a Normalized TAR of 0.31.

\paragraph{Worst-Performing Strategy} The strategy ``G1-P2-I4-C2'' (decentralized governance, selective participation,
selective point-to-point interaction, and self-summarized context) performs poorly, with a Normalized TAR of 0.01.
Despite achieving 50.8\% accuracy, it incurs exorbitant token costs ($\#$Input Token 348,035 and $\#$Output Token 58,795),
highlighting the inefficiency of decentralized systems with fragmented communication.

\subsection{EBFC Task (Structured Evidence Synthesis)}
In the EBFC task, agents must verify factual claims by synthesizing pre-labeled evidence, requiring
persuasion of peers with irrelevant inputs.
The results reveal the following insights:

\paragraph{Optimal Strategy} The strategy ``G2-P3-I1-C3'' (centralized governance, instructor-led participation,
simultaneous interaction, and instructor-curated context summarization) achieves the highest Normalized TAR
of 1.0, with an accuracy of 86.9\% and low token counts ($\#$Input Token 2,111 and $\#$Output Token 490).
This configuration closely matches the theoretical upper bound set by the best-performing individual agent
(88.7\% accuracy) while maintaining computational efficiency.

\paragraph{Decentralized Strategies} Decentralized configurations, such as ``G1-P1-I2-C2'' and ``G1-P2-I4-C2,''
achieve high accuracy (84.4\% and 86.4\%, respectively) but at significantly higher token costs.
For example, ``G1-P2-I4-C2'' achieves 86.4\% accuracy but requires 30,085 input tokens and 6,125 output tokens, r
esulting in a Normalized TAR of 0.07.

\paragraph{Worst-Performing Strategy} The strategy ``G1-P1-I1-C1'' (decentralized governance, full
participation, simultaneous interaction, and full log retention) performs poorly, with a Normalized TAR of 0.06.
It achieves only 49.3\% accuracy while consuming high token costs ($\#$Input Token 28,099 and $\#$Output Token 1,990),
underscoring the inefficiency of decentralized systems with redundant communication.

\subsection{Summary}
\begin{enumerate}
    \item Centralized Governance Dominates: Centralized strategies consistently achieve higher Normalized TAR values across both tasks, demonstrating their ability to balance accuracy and computational efficiency.
    \item Ordered Interaction Patterns: Ordered one-by-one interaction (I2) outperforms simultaneous-talk (I1) and selective point-to-point (I4) patterns, particularly in the PDDP task, where it reduces redundancy and improves token efficiency.
    \item Context Summarization: Summary by the Instructor (C3) significantly reduces token costs compared to full log of the last round (C1) or self-summarized context (C2), especially in the EBFC task.
    \item Task-Specific Dynamics: In the PDDP task, decentralized systems can achieve competitive accuracy but at high computational costs. In contrast, the EBFC task benefits more from centralized control due to the need to filter out irrelevant evidence.
\end{enumerate}

By introducing the TAR and Normalized TAR, we provide a quantitative framework for evaluating multi-agent
collaboration strategies, enabling practitioners to optimize both decision quality and resource utilization.
The results underscore the importance of strategic design choices in scaling LLM-based multi-agent systems for real-world applications.

\begin{table*}[htb]
\centering
\begin{tabular}{c|c|c|c|c|c}
\hline
\textbf{Methods}            & \textbf{Accuracy}$\uparrow$ & \textbf{$\#$Input Token}$\downarrow$ & \textbf{$\#$Output Token}$\downarrow$ & \textbf{Round}$\downarrow$ & \textbf{Normalized TAR}      \\ \hline
$Agent_{\text{all}}$ & 57.8                                    & 1,281                                        & 129                                         & 1.00                                        &  N/A    \\
MV                   & 47.2                                    & 1,873                                        & 661                                         & 5.00                                        &  N/A    \\ \hline
G1-P1-I1-C1          & 50.7                                    & 25,663                                       & 2,184                                        & 3.28                                        & 0.21 \\
G1-P1-I2-C1          & 57.8                                    & \underline{6,470}                           & \underline{854}               & \underline{1.30}              & \underline{0.82} \\
G1-P1-I3-C1          & 45.2                                    & 9,531                                        & 1,127                                        & 1.65                                        & 0.45 \\
G1-P1-I1-C2          & 46.2                                    & 52,400                                       & 15,568                                       & 4.43                                        & 0.06 \\
G1-P1-I2-C2          & \textbf{59.8}                           & 15,057                                       & 3,046                                        & 1.56                                        & 0.31 \\
G1-P1-I3-C2          & 46.7                                    & 19,673                                       & 4,100                                        & 1.81                                        & 0.18 \\
G1-P2-I4-C2          & 50.8                                    & 348,035                                      & 58,795                                       & 9.91                                        & 0.01 \\
G2-P3-I1-C3          & 46.2                                    & 13,119                                       & 2,412                                        & 2.04                                        & 0.28 \\
G2-P3-I2-C3          & \underline{58.8}           & \textbf{4,867}                                        & \textbf{841}                                & \textbf{1.03}                               & \textbf{1}    \\ \hline
\end{tabular}
\caption{Performance of multi-agent collaboration on the PDDP task with different collaboration strategies including Token-Accuracy Ratio.
$Agent_{\text{all}}$: inference by an individual agent with all information concatenated.
    MV: major voting of all agents.
    The best and the second best results are in \textbf{bold} and \underline{underlined}.}
\label{pddp-tar}
\end{table*}

\begin{table*}[htb]
\centering
\begin{tabular}{c|c|c|c|c|c}
\hline
\textbf{Methods}            & \textbf{Accuracy}$\uparrow$ & \textbf{$\#$Input Token}$\downarrow$ & \textbf{$\#$Output Token}$\downarrow$ & \textbf{Round}$\downarrow$ & \textbf{Normalized TAR}      \\ \hline
$Agent_{\text{consistent}}$ & 88.7                                    & 177                                         & 48                                          & 1.00                                        & N/A                 \\ \hline
G1-P1-I1-C1                 & 49.3                                    & 28,099                                       & 1,990                                        & 6.28                                        & 0.06                \\
G1-P1-I2-C1                 & 70.4                                    & 13,361                                       & 1,155                                        & 3.35                                        & 0.18                \\
G1-P1-I3-C1                 & 68.8                                    & 15,368                                       & 1,284                                        & 3.99                                        & 0.16                \\
G1-P1-I1-C2                 & 81.4                                    & 20,074                                       & 6,780                                        & 3.46                                        & 0.08                \\
G1-P1-I2-C2                 & 84.4                                    & 14,600                                       & 3,073                                        & 2.00                                        & 0.15                \\
G1-P1-I3-C2                 & 77.9                                    & 13,125                                       & 2,901                                        & 2.14                                        & 0.15                \\
G1-P2-I4-C2                 & \underline{86.4}                              & 30,085                                       & 6,125                                        & 2.76                                        & 0.07                \\
G2-P3-I1-C3                 & \textbf{86.9}                           & \textbf{2,111}                               & \underline{490}                                   & \underline{1.16}                                  & \textbf{1}          \\
G2-P3-I2-C3                 & 85.4                                    & \underline{2,859}                                  & \textbf{452}                                & \textbf{1.10}                               & \underline{0.86} \\ \hline
\end{tabular}
\caption{Performance of multi-agent collaboration on the EBFC task with different collaboration strategies including Token-Accuracy Ratio.
$Agent_{\text{consistent}}$ is the theoretical upper bound of the multi-agent systems.
The best and the second best results are in \textbf{bold} and \underline{underlined}.}
\label{ebfc-tar}
\end{table*}

%% file: collaboration.bbl
\begin{thebibliography}{17}
\providecommand{\natexlab}[1]{#1}

\bibitem[{Anthropic(2025)}]{Anthropic2025Claude}
Anthropic. 2025.
\newblock \href {https://www.anthropic.com/claude} {Claude: A family of
  language models}.
\newblock Accessed: 2025-02-05.

\bibitem[{Chan et~al.(2024)Chan, Chen, Su, Yu, Xue, Zhang, Fu, and
  Liu}]{chanchateval}
Chi-Min Chan, Weize Chen, Yusheng Su, Jianxuan Yu, Wei Xue, Shanghang Zhang,
  Jie Fu, and Zhiyuan Liu. 2024.
\newblock Chateval: Towards better llm-based evaluators through multi-agent
  debate.
\newblock In \emph{The Twelfth International Conference on Learning
  Representations}.

\bibitem[{Chen et~al.(2024)Chen, Su, Zuo, Yang, Yuan, Chan, Yu, Lu, Hung, Qian
  et~al.}]{chen2023agentverse}
Weize Chen, Yusheng Su, Jingwei Zuo, Cheng Yang, Chenfei Yuan, Chi-Min Chan,
  Heyang Yu, Yaxi Lu, Yi-Hsin Hung, Chen Qian, et~al. 2024.
\newblock Agentverse: Facilitating multi-agent collaboration and exploring
  emergent behaviors.
\newblock In \emph{The Twelfth International Conference on Learning
  Representations}.

\bibitem[{Du et~al.(2024{\natexlab{a}})Du, Li, Torralba, Tenenbaum, and
  Mordatch}]{duimproving}
Yilun Du, Shuang Li, Antonio Torralba, Joshua~B Tenenbaum, and Igor Mordatch.
  2024{\natexlab{a}}.
\newblock Improving factuality and reasoning in language models through
  multiagent debate.
\newblock In \emph{Forty-first International Conference on Machine Learning}.

\bibitem[{Du et~al.(2024{\natexlab{b}})Du, Qian, Liu, Xie, Wang, Dang, Chen,
  and Yang}]{du2024multi}
Zhuoyun Du, Chen Qian, Wei Liu, Zihao Xie, Yifei Wang, Yufan Dang, Weize Chen,
  and Cheng Yang. 2024{\natexlab{b}}.
\newblock Multi-agent software development through cross-team collaboration.
\newblock \emph{arXiv preprint arXiv:2406.08979}.

\bibitem[{Glockner et~al.(2024)Glockner, Stali{\=u}nait{\.e}, Thorne, Vallejo,
  Vlachos, and Gurevych}]{glockner2024ambifc}
Max Glockner, Ieva Stali{\=u}nait{\.e}, James Thorne, Gisela Vallejo, Andreas
  Vlachos, and Iryna Gurevych. 2024.
\newblock Ambifc: Fact-checking ambiguous claims with evidence.
\newblock \emph{Transactions of the Association for Computational Linguistics},
  12:1--18.

\bibitem[{Johnson et~al.(2016)Johnson, Pollard, Shen, Lehman, Feng, Ghassemi,
  Moody, Szolovits, Anthony~Celi, and Mark}]{johnson2016mimic}
Alistair~EW Johnson, Tom~J Pollard, Lu~Shen, Li-wei~H Lehman, Mengling Feng,
  Mohammad Ghassemi, Benjamin Moody, Peter Szolovits, Leo Anthony~Celi, and
  Roger~G Mark. 2016.
\newblock Mimic-iii, a freely accessible critical care database.
\newblock \emph{Scientific data}, 3(1):1--9.

\bibitem[{Kim et~al.(2024)Kim, Park, Jeong, Chan, Xu, McDuff, Lee, Ghassemi,
  Breazeal, and Park}]{kim2024mdagents}
Yubin Kim, Chanwoo Park, Hyewon Jeong, Yik~Siu Chan, Xuhai Xu, Daniel McDuff,
  Hyeonhoon Lee, Marzyeh Ghassemi, Cynthia Breazeal, and Hae~Won Park. 2024.
\newblock Mdagents: An adaptive collaboration of llms for medical
  decision-making.
\newblock In \emph{The Thirty-eighth Annual Conference on Neural Information
  Processing Systems}.

\bibitem[{Li et~al.(2024)Li, Zhang, Yu, FU, and Ye}]{li2024more}
Junyou Li, Qin Zhang, Yangbin Yu, QIANG FU, and Deheng Ye. 2024.
\newblock \href {https://openreview.net/forum?id=bgzUSZ8aeg} {More agents is
  all you need}.
\newblock \emph{Transactions on Machine Learning Research}.

\bibitem[{OpenAI(2022)}]{chatgpt}
OpenAI. 2022.
\newblock Chatgpt.

\bibitem[{Qian et~al.(2025)Qian, Xie, Wang, Liu, Zhu, Xia, Dang, Du, Chen,
  Yang, Liu, and Sun}]{qian2025scaling}
Chen Qian, Zihao Xie, YiFei Wang, Wei Liu, Kunlun Zhu, Hanchen Xia, Yufan Dang,
  Zhuoyun Du, Weize Chen, Cheng Yang, Zhiyuan Liu, and Maosong Sun. 2025.
\newblock \href {https://openreview.net/forum?id=K3n5jPkrU6} {Scaling large
  language model-based multi-agent collaboration}.
\newblock In \emph{The Thirteenth International Conference on Learning
  Representations}.

\bibitem[{Stone and Veloso(2000)}]{stone2000multiagent}
Peter Stone and Manuela Veloso. 2000.
\newblock Multiagent systems: A survey from a machine learning perspective.
\newblock \emph{Autonomous Robots}, 8:345--383.

\bibitem[{Su et~al.(2024)Su, Chen, TANG, Zheng, Li, Yin, Ouyang, and
  Dong}]{su2024two}
Haoyang Su, Renqi Chen, SHIXIANG TANG, Xinzhe Zheng, Jinzhe Li, Zhenfei Yin,
  Wanli Ouyang, and Nanqing Dong. 2024.
\newblock \href {https://openreview.net/forum?id=yYQLvofQ1k} {Two heads are
  better than one: A multi-agent system has the potential to improve scientific
  idea generation}.

\bibitem[{Touvron et~al.(2023)Touvron, Lavril, Izacard, Martinet, Lachaux,
  Lacroix, Rozi{\`e}re, Goyal, Hambro, Azhar et~al.}]{touvron2023llama}
Hugo Touvron, Thibaut Lavril, Gautier Izacard, Xavier Martinet, Marie-Anne
  Lachaux, Timoth{\'e}e Lacroix, Baptiste Rozi{\`e}re, Naman Goyal, Eric
  Hambro, Faisal Azhar, et~al. 2023.
\newblock Llama: Open and efficient foundation language models.
\newblock \emph{arXiv preprint arXiv:2302.13971}.

\bibitem[{Zeng et~al.(2023)Zeng, Liu, Du, Wang, Lai, Ding, Yang, Xu, Zheng,
  Xia, Tam, Ma, Xue, Zhai, Chen, Liu, Zhang, Dong, and Tang}]{zeng2023glm-130b}
Aohan Zeng, Xiao Liu, Zhengxiao Du, Zihan Wang, Hanyu Lai, Ming Ding, Zhuoyi
  Yang, Yifan Xu, Wendi Zheng, Xiao Xia, Weng~Lam Tam, Zixuan Ma, Yufei Xue,
  Jidong Zhai, Wenguang Chen, Zhiyuan Liu, Peng Zhang, Yuxiao Dong, and Jie
  Tang. 2023.
\newblock \href {https://openreview.net/forum?id=-Aw0rrrPUF} {{GLM}-130b: An
  open bilingual pre-trained model}.
\newblock In \emph{The Eleventh International Conference on Learning
  Representations (ICLR)}.

\bibitem[{Zhang et~al.(2024)Zhang, Yue, Sun, Wan, Yu, Fang, Wang, and
  Cheng}]{zhang2024g}
Guibin Zhang, Yanwei Yue, Xiangguo Sun, Guancheng Wan, Miao Yu, Junfeng Fang,
  Kun Wang, and Dawei Cheng. 2024.
\newblock G-designer: Architecting multi-agent communication topologies via
  graph neural networks.
\newblock \emph{arXiv preprint arXiv:2410.11782}.

\bibitem[{Zhuge et~al.(2024)Zhuge, Wang, Kirsch, Faccio, Khizbullin, and
  Schmidhuber}]{zhugegptswarm}
Mingchen Zhuge, Wenyi Wang, Louis Kirsch, Francesco Faccio, Dmitrii Khizbullin,
  and J{\"u}rgen Schmidhuber. 2024.
\newblock Gptswarm: Language agents as optimizable graphs.
\newblock In \emph{Forty-first International Conference on Machine Learning}.

\end{thebibliography}
